\documentclass[aps, prl, twocolumn, superscriptaddress, 10pt]{revtex4-1}

\usepackage{graphicx}
\usepackage{color}
\usepackage{textcomp}

\begin{document}

\title{Extreme field-sensitivity of the magnetic tunneling in Fe-doped Li$_3$N}
\author{M. Fix}
 \affiliation{EP VI, Center for Electronic Correlations and Magnetism, Institute of Physics, University of Augsburg, D-86159 Augsburg, Germany}
\author{J. H. Atkinson}
 \affiliation{Department of Physics, University of Central Florida, Orlando FL 32816, USA}
\author{P. C. Canfield}
 \affiliation{The Ames Laboratory, Iowa State University, Ames, Iowa 50011, USA}
 \affiliation{Department of Physics and Astronomy, Iowa State University, Ames, Iowa 50011, USA}
\author{E. del Barco}
 \affiliation{Department of Physics, University of Central Florida, Orlando FL 32816, USA}
\author{A. Jesche}
 \email[]{anton.jesche@physik.uni-augsburg.de}
 \affiliation{EP VI, Center for Electronic Correlations and Magnetism, Institute of Physics, University of Augsburg, D-86159 Augsburg, Germany}

\begin{abstract}
The magnetic properties of dilute Li$_2$(Li$_{1-x}$Fe$_x$)N with $x \sim 0.001$ are dominated by the spin of single, isolated Fe atoms.
Below $T = 10$\,K the spin-relaxation times become temperature-independent indicating a crossover from thermal excitations to the quantum tunneling regime. 
We report on a strong increase of the spin-flip probability in \textit{transverse} magnetic fields that proves the resonant character of this tunneling process. 
\textit{Longitudinal} fields, on the other hand, lift the ground-state degeneracy and destroy the tunneling condition.
An increase of the relaxation time by four orders of magnitude in applied fields of only a few milliTesla reveals exceptionally sharp tunneling resonances.
Li$_2$(Li$_{1-x}$Fe$_x$)N represents a comparatively simple and clean model system that opens the possibility to study quantum tunneling of the magnetization at liquid helium temperatures. 
\end{abstract}

\maketitle
The understanding of quantum tunneling of the magnetic moment in nanoscale systems experienced a tremendous advance after the discovery of single-molecule magnets (SMMs)\,\cite{Lis1980}, 
with the observation of benchmark effects such as resonant quantum tunneling\,\cite{Sessoli1993, Barbara1995, Friedman1996, Thomas1996} and topological quantum interference\,\cite{Wernsdorfer1999}, followed by numerous studies of the quantum dynamics of the spin in these molecular systems (see\,\cite{Chudnovsky1998} for a review).
Aside of the advance in fundamental understanding, SMMs have been proposed for exciting technological applications, including quantum computation\,\cite{Leuenberger2001,Tejada2001},
magnetic data storage/operation\,\cite{Timm2006},
and magnetic field sensing\,\cite{Leuenberger2006}. 
The latter is based on the unique magnetic field dependence of the tunneling splittings between spin levels in these molecules, which can be finely tuned with small field variations\,\cite{Wernsdorfer1999,Loss1992,Delft1992,delBarco2003,Foss-Feig2009,Atkinson2014}.

Mononuclear SMMs, where the magnetism arises from a single magnetic ion within the molecule, have taken the scene in recent years\,\cite{Ishikawa2003,Alam2006,AlDamen2008,Zadrozny2013,Kazin2014,Jiang2016}.
This is partly due to a decreased number of degrees of freedom in the system due to the absence of exchange-coupled spins, where spin-orbit interaction of an isolated spin with the molecular crystal field governs the magnetic properties of the molecule, giving rise to high local symmetries. 
This results in record-high magnetic anisotropy barriers against magnetization reversal and, consequently, magnetic bistability at high temperatures, or the ability of magnetically dilute single crystals for enhanced quantum coherent dynamics. 

Quantum tunneling has also been observed in inorganic materials with diluted magnetic atoms at subKelvin temperatures, with Ho-doped LiYF$_4$ as a prime example\,\cite{Giraud2001, Barbara2004}. Their similarity to SMMs led these systems to be considered 'single atom magnets'\,\cite{Giraud2003, Bertaina2006}. Only recently, large magnetic anisotropy and magnetic stability were found in a new single-atom magnet based on atomically-doped, insulating bulk system Li$_2$(Li$_{1-x}$Fe$_x$)N\,\cite{Jesche2014b,Fix2017} at temperatures comparable to lanthanide-based mononuclear SMMs, and over two orders of magnitude higher than in previous single-atom magnets.
There is good agreement between the magnetic anisotropy observed experimentally on single crystals (13--27\,meV\,\cite{Jesche2014b,Jesche2015}) and theoretical predictions based on local density approximation\,\cite{Klatyk2002, Novak2002, Antropov2014}, a Green's function method\,\cite{Ke2015} and quantum cluster calculations\,\cite{Xu2017}. 
Strong deviation of the relaxation time from Arrhenius behavior at low temperatures, steps in isothermal $M$-$H$ loops and blocking of the relaxation by applied magnetic fields\,\cite{Jesche2014b} pointed to single-atom magnet behavior.
However, there has not been direct evidence so far for resonant quantum tunneling as the source of the observed magnetic relaxation.

In this letter we report on the effects of transverse and longitudinal magnetic fields on the spin relaxation in Li$_2$(Li$_{1-x}$Fe$_x$)N with small Fe concentrations of $x \sim 0.001$. Isothermal magnetization, ac magnetic susceptibility, and time relaxation studies were performed on single crystals in the temperature range $T = 0.23{-}50$\,K. 
We find not only a clear increase of the relaxation in the presence of transverse magnetic fields, demonstrating quantum tunneling of the magnetization (QTM), but also an extraordinary effect of minuscule magnetic fields on the spin reversal process, where the magnetization relaxation rate can be varied by up to 4 orders of magnitude with field variations of only 2--3\,mT.

\begin{figure}
\includegraphics[width=.45\textwidth]{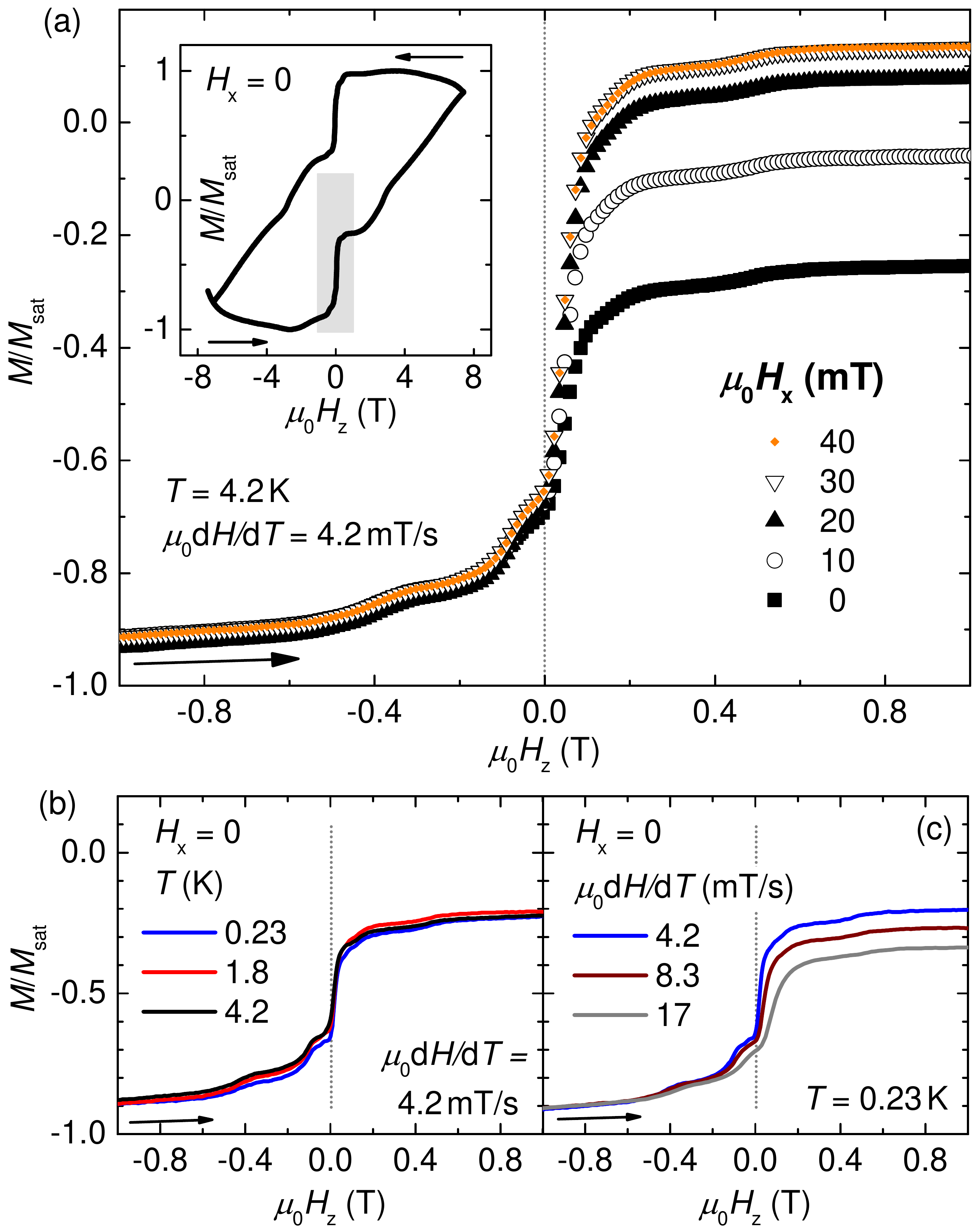}
\caption{(color online) Spin reversal in transverse fields for Li$_2$(Li$_{0.994}$Fe$_{0.006}$)N ($M \parallel H_\mathrm{z} \parallel c, H_\mathrm{x} \perp c$). Shown are sections of isothermal $M$-$H$ half loops recorded after field cooling in $\mu_0H = -7.5$\,T.
(a) A full $M$-$H$ loop is given as inset. The step in $M$ at $H_\mathrm{z} \approx 0$ roughly doubles its size in a transverse field of $\mu_0H_\mathrm{x} = 20$\,mT.
The effect is already saturating by $\mu_0H_\mathrm{x} \approx 30$\,mT.
Compared to the presence of an applied transverse field, the effects of temperature and sweep-rate on the step-size are small as shown in (b) and (c), respectively.} 
\label{hysteresis}
\end{figure}

Single crystals of several millimeters along a side were grown from a lithium-rich flux \,\cite{Jesche2014b, Jesche2014c}.
The Fe concentration $x$ was inferred from the known saturation magnetization\,\cite{Jesche2014b} and is close to the nominal one.
For the basic characterization of the quantum tunneling behavior of this system at low temperature, a $\sim$1$\times$0.2$\times$0.2\,mm$^3$ single crystal of Li$_2$(Li$_{0.994}$Fe$_{0.006}$)N was placed directly on top of the sensing area of a high-sensitivity 2-dimensional electron gas micro-Hall effect magnetometer placed on an Oxford Instruments Heliox $^3$He cryostat ($T_{\rm base} = 230$\,mK). 
A three-dimensional superconducting vector magnet was employed to apply magnetic fields in arbitrary directions with respect to the crystallographic c-axes of the sample, with a maximum longitudinal field of 8 Tesla.
Figure\,\ref{hysteresis}a shows sections of $\mu_0H_\mathrm{z}$-hysteresis loops (-8\,T\,{-}\,0\,{-}\,+8\,T) obtained at $T = 4.2$\,K in transverse fields of $\mu_0H_\mathrm{x} =$ 0--40\,mT.
The longitudinal field was applied along the easy axis, that is $H \parallel c$, and swept at a rate of $\mu_0\mathrm{d}H_\mathrm{z}/\mathrm{d}t = $4.2\,mT/s (a full loop is given in the inset; The gray area marks the section plotted in the main panel).
Several magnetization jumps are observed similar to the earlier report\,\cite{Jesche2014b}: aside of the main step at zero field additional smaller ones appear at $\mu_0H = 0.12, 0.45$, and 2.7\,T. 
In this work we focus on the main jump at $H \approx 0$ and the spin reversal in small fields $\mu_0H_\mathrm{z} < 10$\,mT which is significantly below the field of the first additional step.

Figure\,\ref{hysteresis}a shows the behavior of the low field region of the hysteretic magnetization curve as a function of a magnetic field applied transverse to the easy anisotropy axis of the sample. 
The zero-field jump quickly increases upon application of a small transverse field ($\mu_0H_\mathrm{x} = 10$\,mT), saturating by $\mu_0 H_\mathrm{x} \sim 30$\,mT\,\cite{SI}. 
Note that the sample magnetization has not yet reached its overall positive saturation value (see Fig.\,\ref{hysteresis}a inset), which is achieved for much larger magnetic field values ($\mu_0H_\mathrm{x} > 8$\,T). 
The saturation of the zero-field jump is due to a narrow quantum tunneling resonance, with spins tunneling at zero longitudinal field towards a local zero magnetization saturation.
When the longitudinal field increases, the system gets out of resonance and spins cannot tunnel any longer, with the magnetization remaining near $M=0$ until phonons equilibrate the system at higher longitudinal fields, as indicated by the subsequent monotonous increase of $M$ vs. $H_\mathrm{z}$. 
The rapid change in the tunneling rate with the magnitude of the applied transverse field makes this system distinct from SMMs, where usually transverse fields at least one order of magnitude larger (e.g., $<0.5$\,T for Mn12\,\cite{delBarco2003}) are necessary to achieve magnetization saturation at a QTM resonance.
One can estimate a ground tunneling splitting of $\Delta_{k=0} = 0.1$\,neV from the change in magnetization in the zero field jump using the Landau-Zener-Stueckelberg formalism for tunneling\,\cite{Zener1932}:

\begin{equation}
\label{Eq:LZS}P_k = 1-\exp[\frac{\pi \Delta_k^2}{2\nu_0\alpha}]\,,
\end{equation}

where $\nu_0=g\mu_B(2J-k)$, being $g$ the magnetic Land\'e factor, $\mu_{\rm B}$ the Bohr magneton, $J$ the spin, $k$ the resonance number, and $\alpha$ the rate at which the magnetic field is swept across the resonance. This value should be taken as an upper-bound estimate, since reshuffling fields may cause the system to cross the resonance multiple times during the field sweep, particularly at low sweep rates. Similar $M$-$H$ loops were recorded for temperatures as low as $T = 0.23$\,K and revealed no appreciable temperature-dependence (Fig.\,\ref{hysteresis}b).
Increasing the sweep-rate by a factor of four also causes weaker changes in $M$-$H$ than those caused by $\mu_0H_\mathrm x = 20$\,mT (Fig.\,\ref{hysteresis}c).
Unfortunately, experiments where the longitudinal field is swept while a fixed transverse field is applied along different directions within the (hard) $a$-$b$ plane\,\cite{SI} do not display any measurable angular modulation and, consequently, in-plane details of the local site symmetry of this species cannot be determined within the resolution of our measurements.
 
\begin{figure}
\includegraphics[width=.45\textwidth]{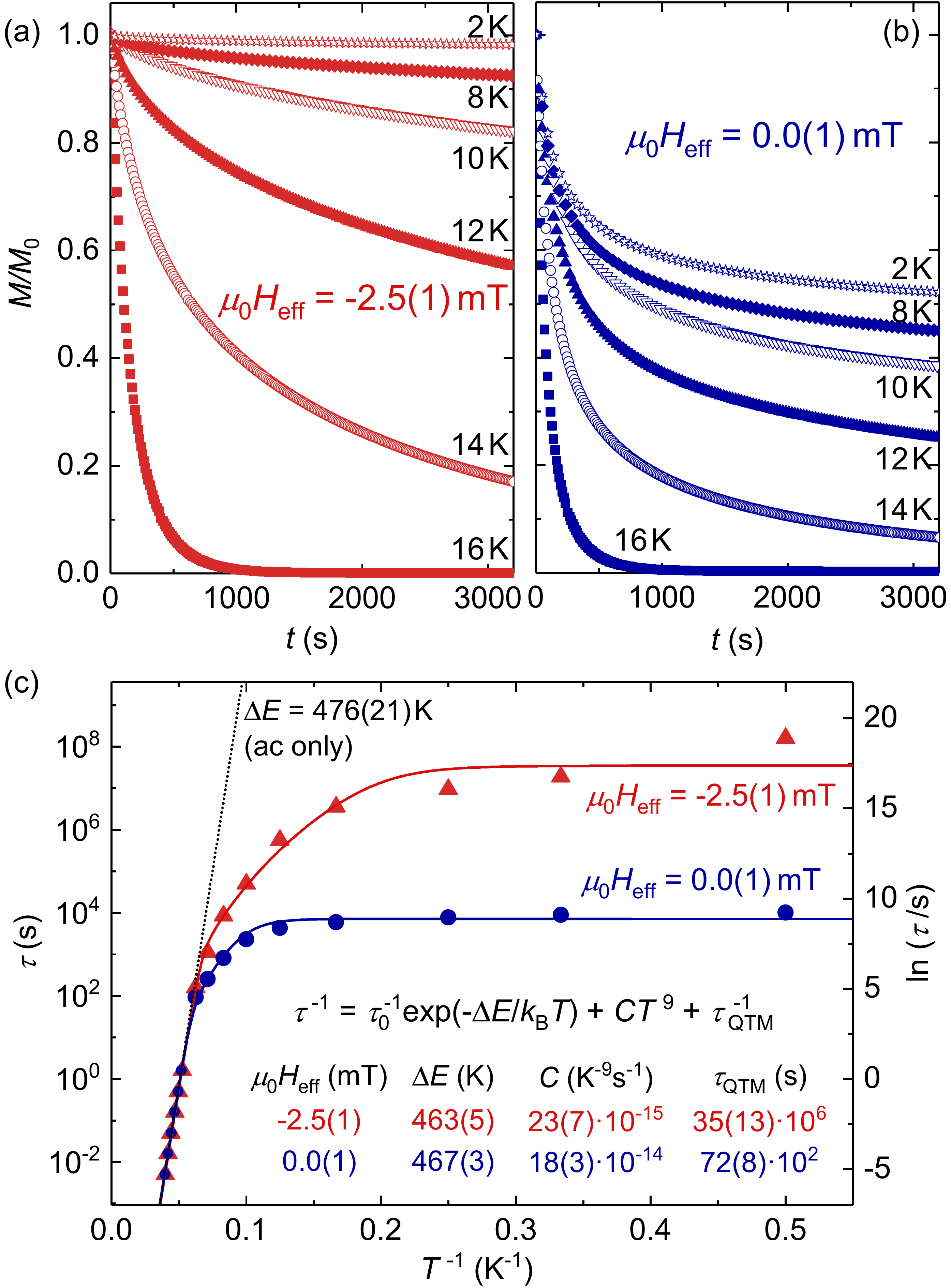}
\caption{(color online) 
Spin reversal in Li$_2$(Li$_{0.999}$Fe$_{0.001}$)N in $H \approx 0$ ($M \parallel c$).
(a) Time-dependence of the magnetization after ramping the applied field from $\mu_0 H = 7$\,T to nominal zero. 
The presence of the longitudinal remnant field of the magnet, applied opposite to the initial magnetization ($\mu_0H_{\rm eff} = -2.5$\,mT), was found to stabilize the orientation of the magnetic moments drastically.
(b) Ramping the field to nominal $\mu_0H = +2.5$\,mT compensates for the remnant field\,\cite{SI} and leads to faster decrease of $M$. 
(c) Relaxation times $\tau$ were determined by fitting a stretched exponential function to $M(t)$ and are shown in form of an Arrhenius plot ($T \leq 16$\,K). 
The solid lines are fits to the equation given in the plot.
For $T > 20$\,K ($\tau \leq 10$\,s) $\tau$ was determined from ac susceptibility and reveals thermally activated behavior (dotted line). 
}
\label{FC}
\end{figure}

To further investigate the nature of quantum tunneling of the isolated Fe centers in this system, magnetic relaxation experiments have been performed on Li$_2$(Li$_{1-x}$Fe$_{x}$)N single crystals with an even lower Fe concentration of $x = 0.001$.
Measuring the decrease of the magnetization from saturation after an applied field is removed has been widely used in order to study spin reversal in various SMM systems. 
The temperature-independence of the spin relaxation - a hallmark of quantum tunneling - is directly observable in the time-dependence of the magnetization\,\cite{Sangregorio1997}. 
For this purpose, a Magnetic Property Measurement System (MPMS3) manufactured by Quantum Design has been employed.
Figure\,\ref{FC}a shows the magnetization of Li$_2$(Li$_{0.999}$Fe$_{0.001}$)N along the crystallographic c-direction as a function of time in (nominal) zero-field (see discussions below). 
Prior to the measurement, the sample was cooled in an applied magnetic field of $\mu_0H = 7$\,T ($H \parallel c$) to the temperature given, followed by ramping the field to zero with a rate of 70\,mT/s ($t = 0$ is defined by reaching $H \approx 0$).
After 1\,h the magnetization remains at 98\% of the initial value at $T = 2$\,K. 
The relaxation changes only slightly for increasing the temperatures to $T = 8$\,K. 
For $T \geq 10$\,K, however, the magnetization decays at significantly higher rates and vanishes at $T = 16$\,K already after $\sim 1/2$\,h.
When the field-dependence of the relaxation process was investigated we found that even the comparatively small remnant field of the magnet has an extreme effect on the relaxation. 
Measurements on superconducting indium reproducibly revealed a remnant field of $\mu_0H_{\rm eff} = -2.5(1)$\,mT after the magnet was set from $\mu_0H = +7$\,T to nominal zero.
Compensating the remnant field by setting the applied field to nominal $\mu_0H = +2.5$\,mT - in accordance with the measurement on In (see \cite{SI}) - leads to significantly enhanced relaxation rates as shown in Fig.\,\ref{FC}b ($H_{\rm eff} = 0$).
Now the magnetization at $T = 2$\,K decreases within 1\,h to roughly 50\,\% of its initial value

%The inset shows a direct comparison of $M(t)$ in $\mu_0H = -2.5$\,mT and $H = 0$.

The relaxation times $\tau$ were determined by fitting a stretched exponential function to the time-dependent magnetization:
\begin{equation}
M(t) = M_{\rm eq} + [M(0) - M_{\rm eq}]\exp\{-(t/\tau)^\beta\}.
\label{stretched}
\end{equation}
With $M_{\rm eq}$ set to zero, 3 free parameters remain (see\,\cite{SI} for $\beta$ and further details).
The obtained values for $\tau$ are depicted in Fig.\,\ref{FC}c in form of an Arrhenius plot.
In $H_{\rm eff} = 0$ the relaxation time for $T < 10$\,K is reduced by up to 4 orders of magnitude compared to $\mu_0H_{\rm eff} = -2.5(1)$\,mT, an effect that is most likely a direct consequence of destroying the tunneling condition by lifting the degeneracy of the ground state doublet\,\cite{Xu2017}. 
A pronounced field-dependence had been also observed in samples with larger Fe concentration\,\cite{Jesche2014b}, however, the crucial effect of the remnant field of the magnet was not taken into account in \cite{Jesche2014b} and the extreme influence of sub-mT fields (see below) remained elusive at that time.
Note that not only the absolute values of $\tau$ but also its temperature-dependence for $T < 10$ is significantly reduced and indicates the 
%negligible role/
irrelevance of thermal fluctuations for spin-reversal under these conditions.

An increase of $\tau$ in applied longitudinal fields is expected in the (thermally assisted) tunneling regime and has been observed in several systems. 
However, we are not aware of any material that
%  even begins to compare 
demonstrates a field-sensitivity comparable to the one found for Li$_2$(Li$_{1-x}$Fe$_x$)N.
The well investigated SMM Fe$_8$, for example, does show a change of $\tau$ by a factor of $\sim10^4$\,\cite{Sangregorio1997}, however, that required a field of $\mu_0H = -100$\,mT and cooling the sample to $T < 400$\,mK.
Note that the field-dependence at $T = 100$\,mK shown in Fig.\,4 of Ref.\,\cite{Sangregorio1997} is weaker than the one shown in Fig.\,\ref{FC}.  
A longitudinal applied field in the order of $\sim\,0.1$\,T typically causes an increase of $\tau$ by a factor of 1--10,\,e.g.\,\cite{Thomas1996,Friedman1998,Fort1998,Fominaya1999,Freedman2010,Fataftah2014}.

For higher temperatures $T > 16$\,K alternating-current susceptibility measurements were employed to determine $\tau$.
The observed Arrhenius behavior ($\tau_{\rm Orbach} = \tau_0\exp\{\Delta E/T \}$, dotted line in Fig.\,\ref{FC}c) indicates thermally activated relaxation that is driven by an Orbach process\,\cite{Orbach1961} in stark contrast to the temperature-independent behavior for $T < 10$\,K.
An effective energy barrier of $\Delta E = 476(21)$\,K and a pre-exponential factor of $\tau_0 \approx 5 \times 10^{-11}$\,s were found for the Arrhenius regime.
Similar and even larger effective energy barriers were demonstrated in other mononuclear SMMs, e.g. $\Delta E = 331$\,K in the lanthanide double-decker [Pc$_2$Tb]$^-\cdot$TBA$^+$\,\cite{Ishikawa2003}, $\Delta E = 469$\,K in an Fe(I) complex\,\cite{Zadrozny2013} or $\Delta E > 1000$\,K in a Dy(III) based SMM\,\cite{Jiang2016}.
Like Li$_2$(Li$_{1-x}$Fe$_x$)N, these compounds show the crossover from thermally activated behavior to quantum tunneling at temperatures of $T_{\rm cr} \sim 10$\,K%indicated by significant deviations from Arrhenius behavior
, however, with $\tau$ being in the range of seconds in contrast to $\tau \sim 10^4$ seconds found for Li$_2$(Li$_{1-x}$Fe$_x$)N.

We associate the small slope of $\tau(T)$ observed at low temperatures, when the field is applied (red, solid triangles in Fig.\,\ref{FC}c), to contributions from Raman processes which give rise to $\tau_{\rm Raman}^{-1} = C T^n$ with $n = 9$ for a doublet ground state\,\cite{Carlin1986}. The overall temperature-dependence of $\tau$ is well described over 10 orders of magnitude by a sum of Orbach, Raman, and quantum tunneling contributions: $\tau^{-1} = \tau_{\rm Orbach}^{-1} +  \tau_{\rm Raman}^{-1} + \tau_{\rm QT}^{-1}$ (solid lines in Fig.\,\ref{FC}c). 
For the sake of simplicity and in order to keep the number of free parameters low, we kept $\tau_0 = 5 \times 10^{-11}$\,s as well as the the exponent $n = 9$ fix and neglect further possible contributions from direct processes ($\tau = A H^2 T$\,\cite{Carlin1986}).
The obtained values are summarized in Fig.\,\ref{FC}c. 
Note that for $\mu_0H_{\rm eff} = -2.5$\,mT the exponent converges to $n = 9$ even when very different starting values are chosen. 
For $H_{\rm} = 0$ the Raman process is efficient only in a small temperature window $10\,{\rm K} < T < 15$\,K and the relaxation dominated by Orbach processes ($T > 15 $\,K) and quantum tunneling ($T < 10$\,K).

%and assume that the field-dependence is fully dominated by the tunneling contribution. }

%excited states (i.e., thermally excited quantum tunneling), which become negligible when the field is eliminated and pure ground state tunneling becomes the dominant relaxation path (hence the plateau).

\begin{figure}
\includegraphics[width=.45\textwidth]{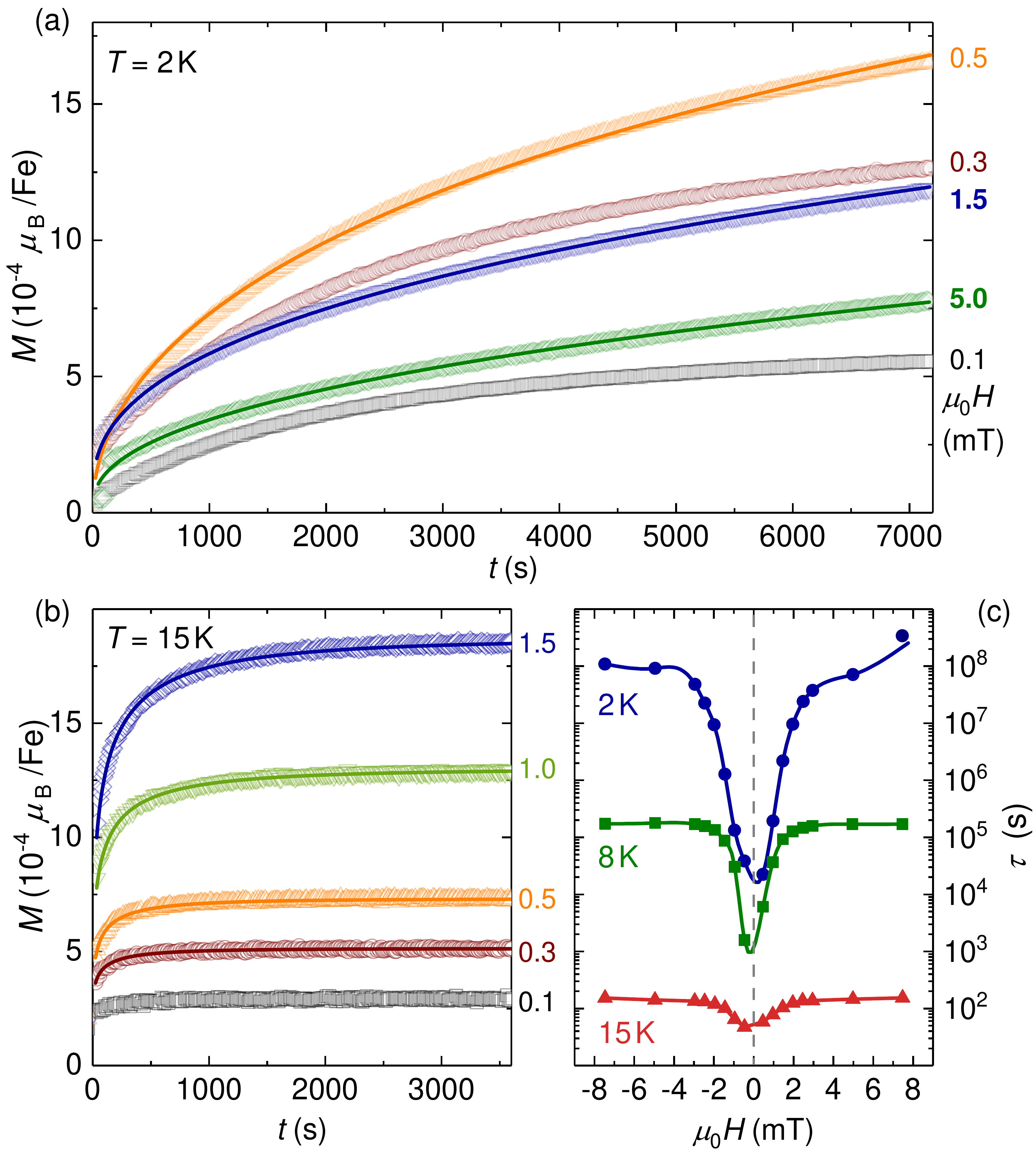}
\caption{(color online)
Time-dependent magnetization of Li$_2$(Li$_{0.999}$Fe$_{0.001}$)N as response to a longitudinal applied field after zero-field cooling ($H \parallel M \parallel c$). 
(a) Increasing the applied field to above $\sim 0.5$\,mT leads to decreasing magnetization in the tunneling regime ($T = 2$\,K). The weaker change in $M(t)$ despite the stronger cause is a consequence of lifting the ground state degeneracy.
(b) At higher temperatures ($T = 15\,$K) the magnetization shows the expected monotonic increase with $H$. The lines are fits to a stretched exponential function. 
(c) Field-dependent relaxation times $\tau(H)$. 
At $T = 2$\,K a small field of $\mu_0H \approx 3$\,mT changes $\tau$ by 4 orders of magnitude. The field-sensitivity decreases with increasing temperature (lines are guides to the eye) \label{ZFC}}
\end{figure}

Analyzing the decay of the magnetization after field-removal in order to extract $\tau = \tau(T,H)$ as performed in the previous section could suffer from the finite time required to ramp the field to zero as well as from significant changes of internal fields during the process\,\cite{Friedman1998}. 
Therefore, we have applied a second method to determine $\tau$ based on the \textit{increase} of $M(t)$ in response to applying small fields ($\mu_0 H < 10$\,mT).
Prior to the measurement the sample was cooled in zero-field ($H_{\rm eff} \approx 0$ was ensured by quenching the magnet before the first run and oscillating $H$ to zero after every following one).
Figure\,\ref{ZFC}a,b show the time-dependent magnetization obtained for Li$_2$(Li$_{0.999}$Fe$_{0.001}$)N at $T = 2$\,K and $T = 15$\,K, respectively, for $H \parallel  M \parallel c$.
%The field was applied parallel and anti-parallel to $c$ in order detect possible effects of a small remnant field. 
Applying the field anti-parallel to $c$ reveals basically symmetric $M(t)$ curves for $\vert \mu_0H \vert > 0.3$\,mT (see\,\cite{SI}) and indicates well-defined $H$ values close to the nominal ones given on the right-hand side of Fig.\,\ref{ZFC}a,b.
Whereas $M(t)$ at $T = 15$\,K shows the expected monotonic increase with $H$, we observe a remarkable field-dependence at $T = 2$\,K:
$M(t)$ increases stronger in smaller $H$!
In fact, $M(t)$ in $\mu_0 H = 0.5$\,mT remains (within the time of the experiment) significantly larger than in $\mu_0 H = 1.5$ and $5.0$\,mT even though the driving force is lower. 
As discussed below, it is the destruction of resonant tunneling by minute longitudinal fields that causes this counter-intuitive behavior. 

Again, we employed eq.\ref{stretched} to estimate $\tau$. In order to keep the number of free parameters low, $M_{\rm eq}$ was calculated assuming a two-level system with $\mu_z = \pm 5\,\mu_{\rm B}$\,\cite{Jesche2014b,Xu2017} and $M_{\rm eq} = 5\mu_{\rm B}\,{\rm tanh}[(5\mu_{\rm B}B)/(k_{\rm B}T)]$. With $M(0)$ fixed to zero and an offset to account for small systematic shifts there are 3 free parameters left (see\,\cite{SI} for further details).
The relaxation times obtained for $T = 2,\,8$\, and 15\,K are plotted in Fig.\,\ref{ZFC}c in a semi-log plot. 
In accordance with the field-cooled decay-measurements presented in the previous section, $\tau$ decreases by four orders of magnitude already in small applied fields of $\mu_0 H \approx 3$\,mT at $T = 2$\,K.
For larger $T$ the field-dependence decreases, however, a clear peak-like anomaly remains in $\tau(H)$ even at $T = 15$\,K.

For the archetypal Mn$_{12}$-acetate, relaxation times were estimated by a similar approach.
The data presented in Refs.\,\cite{Friedman1998,Zhong2000} indicate that an applied field in the range of $\mu_0H \approx 3$\,mT has no measurable effect on the spin reversal [although a \textit{single} exponential fit that was restricted to the long-time tail ($t > 2000$\,s) was performed in order to estimate the relaxation rate].

So, why is the field-dependence of $\tau$ in Li$_2$(Li$_{1-x}$Fe$_x$)N by orders of magnitude larger than in SMMs despite similar energy barriers and crossover temperatures? And most importantly, why does this single-atom magnet based on a transition metal ion present magnetic bistability at temperatures two orders of magnitude larger than those of previously reported single-atom magnets based on rare-earths, with substantially larger spin-orbit coupling? 
Although we cannot provide a definite answer to either of these two questions with the data in hand, we can provide potential scenarios addressing the former according to the experimental observations in this work. 
Further experimentation and modeling will be needed to address the latter, in particular since an appropriate, effective spin Hamiltonian was not formulated so far. 
The main difficulties are given by the presence of unquenched orbital moments (that give rise to the large magnetic anisotropy\,\cite{Klatyk2002,Novak2002,Jesche2014b}), their coupling to the spin (characterizing the ground state by four doublets that are best characterized by the quantum numbers $m_J = \{\pm 7/2, \pm 5/2, \pm 3/2, \pm 1/2 \}$\,\cite{Zadrozny2013,Xu2017} appears reasonable but has not been rigorously proven) and the presence of Fe 3d-4s hybridization\,\cite{Klatyk2002,Novak2002}.
These characteristics make this system particularly special, since they provide a novel source of hysteretic behavior distinct from those purely coming from spin-orbit coupling in other $d$- and $f$-electron systems, as is the case of other SMMs and the Ho-based single atom magnets reported before. 

Qualitatively, a single Fe center embedded in the insulating Li$_3$N matrix is more isolated (average Fe-Fe distance ${\sim}\,36$\,\r{A}) and less coupled to any magnetic or non-magnetic degrees of freedom than the magnetic centers in cluster SMM systems.
The Fe-Fe dipolar coupling at average distance amounts to only ${\sim}\,0.06$\,\textmu eV (along the c-axis, and half of this value in-plane). This is substantially smaller than the typical dipolar field values found in condensed crystals of SMMs (e.g. ${\sim}\,20{-}30$\,\textmu eV in Fe$_8$ or Mn$_{12}$\,\cite{McHugh2009}), and still smaller than the average dipolar broadening (${\sim}\,2{-}3$\,\textmu eV\,\cite{Giraud2001}) observed in Ho-based single-atom magnets with similar dilution concentrations. 
Furthermore, the defect concentration in the rather simple binary Li$_3$N is expected to be lower than in solids that are build from large organic molecules, and as such the dispersion of transverse terms caused by dislocation-induced strain\,\cite{Chudnovsky2001} results less critical in our system. 
In particular, a smaller presence of defects in our crystals would minimize tilts of the ions' easy axis (that gives rise to varying internal fields). 
Accordingly, the distribution of relaxation times is less broad and the tunneling resonance sharper.
It remains to be seen if this is also reflected in large coherence times that allowed for the observation of Rabi oscillations in other single-atom magnets\,\cite{Bertaina2009, Rakhmatullin2009} and are essential for potential applications in quantum computing\,\cite{Bertaina2007}.

It is therefore the narrow width of the resonance what causes the strong observed field dependence. 
Given that 3\,mT are sufficient to lift the zero-field degeneracy by Zeeman splitting, as indicated by the peak width in Fig.\,\ref{ZFC}c, we obtain an energy width of 1.7\,\textmu eV for the ground state (which is assumed an effective $J = \pm 7/2$ doublet with $\mu_z = \pm 5\,\mu_{\rm B}$\,\cite{Xu2017}).
The smaller magnetization in larger applied fields is a direct consequence of destroying this sharp resonance condition and manifests a unique example for a 'larger cause but smaller effect scenario'.

To summarize, Fe-doped Li$_3$N allows the study of resonant quantum tunneling of the magnetization in a comparatively simple and clean system.
The Fe-atoms behave like a SMM and can be considered a single-atom magnet.
The marked monodispersity degree and the ability to tune the concentration of spins places this system as an ideal candidate to study the quantum dynamics of anisotropic spins. 
This, together with the strong field-dependence of the spin-reversal allows to create stable ($\mu_0H = 3$\,mT) but switchable ($H = 0$) states that could act as a 'quantum bit' at elevated temperatures.

\begin{acknowledgments}
We thank Michael L. Baker, Jonathan R. Friedman, Liviu Hozoi and Hans-Henning Klauss for fruitful discussions. This work was supported by the Deutsche Forschungsgemeinschaft (DFG, German Research Foundation) - Grant No. JE 748/1 and the U.S. Department of Energy, Office of Basic Energy Science, Division of Materials Sciences and Engineering; part of the research was performed at the Ames Laboratory. Ames Laboratory is operated for the U.S. Department of Energy by Iowa State University [contract number DE-AC02-07CH11358]. J.H.A. and E.d.B. acknowledge support from the U. S. National Science Foundation under Grants No. DMR-1503627 and No. DMR-1630174.
\end{acknowledgments}


\begin{thebibliography}{51}%
\makeatletter
\providecommand \@ifxundefined [1]{%
 \@ifx{#1\undefined}
}%
\providecommand \@ifnum [1]{%
 \ifnum #1\expandafter \@firstoftwo
 \else \expandafter \@secondoftwo
 \fi
}%
\providecommand \@ifx [1]{%
 \ifx #1\expandafter \@firstoftwo
 \else \expandafter \@secondoftwo
 \fi
}%
\providecommand \natexlab [1]{#1}%
\providecommand \enquote  [1]{``#1''}%
\providecommand \bibnamefont  [1]{#1}%
\providecommand \bibfnamefont [1]{#1}%
\providecommand \citenamefont [1]{#1}%
\providecommand \href@noop [0]{\@secondoftwo}%
\providecommand \href [0]{\begingroup \@sanitize@url \@href}%
\providecommand \@href[1]{\@@startlink{#1}\@@href}%
\providecommand \@@href[1]{\endgroup#1\@@endlink}%
\providecommand \@sanitize@url [0]{\catcode `\\12\catcode `\$12\catcode
  `\&12\catcode `\#12\catcode `\^12\catcode `\_12\catcode `\%12\relax}%
\providecommand \@@startlink[1]{}%
\providecommand \@@endlink[0]{}%
\providecommand \url  [0]{\begingroup\@sanitize@url \@url }%
\providecommand \@url [1]{\endgroup\@href {#1}{\urlprefix }}%
\providecommand \urlprefix  [0]{URL }%
\providecommand \Eprint [0]{\href }%
\providecommand \doibase [0]{http://dx.doi.org/}%
\providecommand \selectlanguage [0]{\@gobble}%
\providecommand \bibinfo  [0]{\@secondoftwo}%
\providecommand \bibfield  [0]{\@secondoftwo}%
\providecommand \translation [1]{[#1]}%
\providecommand \BibitemOpen [0]{}%
\providecommand \bibitemStop [0]{}%
\providecommand \bibitemNoStop [0]{.\EOS\space}%
\providecommand \EOS [0]{\spacefactor3000\relax}%
\providecommand \BibitemShut  [1]{\csname bibitem#1\endcsname}%
\let\auto@bib@innerbib\@empty
%</preamble>
\bibitem [{\citenamefont {Lis}(1980)}]{Lis1980}%
  \BibitemOpen
  \bibfield  {author} {\bibinfo {author} {\bibfnamefont {T.}~\bibnamefont
  {Lis}},\ }\href {\doibase 10.1107/S0567740880007893} {\bibfield  {journal}
  {\bibinfo  {journal} {Acta Crystallographica Section B}\ }\textbf {\bibinfo
  {volume} {36}},\ \bibinfo {pages} {2042} (\bibinfo {year}
  {1980})}\BibitemShut {NoStop}%
\bibitem [{\citenamefont {Sessoli}\ \emph {et~al.}(1993)\citenamefont
  {Sessoli}, \citenamefont {Gatteschi}, \citenamefont {Caneschi},\ and\
  \citenamefont {Novak}}]{Sessoli1993}%
  \BibitemOpen
  \bibfield  {author} {\bibinfo {author} {\bibfnamefont {R.}~\bibnamefont
  {Sessoli}}, \bibinfo {author} {\bibfnamefont {D.}~\bibnamefont {Gatteschi}},
  \bibinfo {author} {\bibfnamefont {A.}~\bibnamefont {Caneschi}}, \ and\
  \bibinfo {author} {\bibfnamefont {M.~A.}\ \bibnamefont {Novak}},\ }\href
  {http://dx.doi.org/10.1038/365141a0} {\bibfield  {journal} {\bibinfo
  {journal} {Nature}\ }\textbf {\bibinfo {volume} {365}},\ \bibinfo {pages}
  {141} (\bibinfo {year} {1993})}\BibitemShut {NoStop}%
\bibitem [{\citenamefont {Barbara}\ \emph {et~al.}(1995)\citenamefont
  {Barbara}, \citenamefont {Wernsdorfer}, \citenamefont {Sampaio},
  \citenamefont {Park}, \citenamefont {Paulsen}, \citenamefont {Novak},
  \citenamefont {Ferr\'e}, \citenamefont {Mailly}, \citenamefont {Sessoli},
  \citenamefont {Caneschi}, \citenamefont {Hasselbach}, \citenamefont
  {Benoit},\ and\ \citenamefont {Thomas}}]{Barbara1995}%
  \BibitemOpen
  \bibfield  {author} {\bibinfo {author} {\bibfnamefont {B.}~\bibnamefont
  {Barbara}}, \bibinfo {author} {\bibfnamefont {W.}~\bibnamefont
  {Wernsdorfer}}, \bibinfo {author} {\bibfnamefont {L.}~\bibnamefont
  {Sampaio}}, \bibinfo {author} {\bibfnamefont {J.}~\bibnamefont {Park}},
  \bibinfo {author} {\bibfnamefont {C.}~\bibnamefont {Paulsen}}, \bibinfo
  {author} {\bibfnamefont {M.}~\bibnamefont {Novak}}, \bibinfo {author}
  {\bibfnamefont {R.}~\bibnamefont {Ferr\'e}}, \bibinfo {author} {\bibfnamefont
  {D.}~\bibnamefont {Mailly}}, \bibinfo {author} {\bibfnamefont
  {R.}~\bibnamefont {Sessoli}}, \bibinfo {author} {\bibfnamefont
  {A.}~\bibnamefont {Caneschi}}, \bibinfo {author} {\bibfnamefont
  {K.}~\bibnamefont {Hasselbach}}, \bibinfo {author} {\bibfnamefont
  {A.}~\bibnamefont {Benoit}}, \ and\ \bibinfo {author} {\bibfnamefont
  {L.}~\bibnamefont {Thomas}},\ }\href {\doibase
  http://dx.doi.org/10.1016/0304-8853(94)00585-0} {\bibfield  {journal}
  {\bibinfo  {journal} {J. Magn. Magn. Mater.}\ }\textbf {\bibinfo {volume}
  {140}},\ \bibinfo {pages} {1825} (\bibinfo {year} {1995})}\BibitemShut
  {NoStop}%
\bibitem [{\citenamefont {Friedman}\ \emph {et~al.}(1996)\citenamefont
  {Friedman}, \citenamefont {Sarachik}, \citenamefont {Tejada},\ and\
  \citenamefont {Ziolo}}]{Friedman1996}%
  \BibitemOpen
  \bibfield  {author} {\bibinfo {author} {\bibfnamefont {J.~R.}\ \bibnamefont
  {Friedman}}, \bibinfo {author} {\bibfnamefont {M.~P.}\ \bibnamefont
  {Sarachik}}, \bibinfo {author} {\bibfnamefont {J.}~\bibnamefont {Tejada}}, \
  and\ \bibinfo {author} {\bibfnamefont {R.}~\bibnamefont {Ziolo}},\ }\href
  {\doibase 10.1103/PhysRevLett.76.3830} {\bibfield  {journal} {\bibinfo
  {journal} {Phys. Rev. Lett.}\ }\textbf {\bibinfo {volume} {76}},\ \bibinfo
  {pages} {3830} (\bibinfo {year} {1996})}\BibitemShut {NoStop}%
\bibitem [{\citenamefont {Thomas}\ \emph {et~al.}(1996)\citenamefont {Thomas},
  \citenamefont {Lionti}, \citenamefont {Ballou}, \citenamefont {Gatteschi},
  \citenamefont {Sessoli},\ and\ \citenamefont {Barbara}}]{Thomas1996}%
  \BibitemOpen
  \bibfield  {author} {\bibinfo {author} {\bibfnamefont {L.}~\bibnamefont
  {Thomas}}, \bibinfo {author} {\bibfnamefont {F.}~\bibnamefont {Lionti}},
  \bibinfo {author} {\bibfnamefont {R.}~\bibnamefont {Ballou}}, \bibinfo
  {author} {\bibfnamefont {D.}~\bibnamefont {Gatteschi}}, \bibinfo {author}
  {\bibfnamefont {R.}~\bibnamefont {Sessoli}}, \ and\ \bibinfo {author}
  {\bibfnamefont {B.}~\bibnamefont {Barbara}},\ }\href {\doibase
  10.1038/383145a0} {\bibfield  {journal} {\bibinfo  {journal} {Nature}\
  }\textbf {\bibinfo {volume} {383}},\ \bibinfo {pages} {145} (\bibinfo {year}
  {1996})}\BibitemShut {NoStop}%
\bibitem [{\citenamefont {Wernsdorfer}\ and\ \citenamefont
  {Sessoli}(1999)}]{Wernsdorfer1999}%
  \BibitemOpen
  \bibfield  {author} {\bibinfo {author} {\bibfnamefont {W.}~\bibnamefont
  {Wernsdorfer}}\ and\ \bibinfo {author} {\bibfnamefont {R.}~\bibnamefont
  {Sessoli}},\ }\href {\doibase 10.1126/science.284.5411.133} {\bibfield
  {journal} {\bibinfo  {journal} {Science}\ }\textbf {\bibinfo {volume}
  {284}},\ \bibinfo {pages} {133} (\bibinfo {year} {1999})}\BibitemShut
  {NoStop}%
\bibitem [{\citenamefont {Chudnovsky}\ and\ \citenamefont
  {Tejada}(1998)}]{Chudnovsky1998}%
  \BibitemOpen
  \bibfield  {author} {\bibinfo {author} {\bibfnamefont {E.~M.}\ \bibnamefont
  {Chudnovsky}}\ and\ \bibinfo {author} {\bibfnamefont {J.}~\bibnamefont
  {Tejada}},\ }\href@noop {} {\emph {\bibinfo {title} {Macroscopic Quantum
  Tunneling of Magnetic Moment}}}\ (\bibinfo  {publisher} {Cambridge University
  Press, Cambridge, England},\ \bibinfo {year} {1998})\BibitemShut {NoStop}%
\bibitem [{\citenamefont {Leuenberger}\ and\ \citenamefont
  {Loss}(2001)}]{Leuenberger2001}%
  \BibitemOpen
  \bibfield  {author} {\bibinfo {author} {\bibfnamefont {M.~N.}\ \bibnamefont
  {Leuenberger}}\ and\ \bibinfo {author} {\bibfnamefont {D.}~\bibnamefont
  {Loss}},\ }\href {\doibase 10.1038/35071024} {\bibfield  {journal} {\bibinfo
  {journal} {Nature}\ }\textbf {\bibinfo {volume} {410}},\ \bibinfo {pages}
  {789} (\bibinfo {year} {2001})}\BibitemShut {NoStop}%
\bibitem [{\citenamefont {Tejada}\ \emph {et~al.}(2001)\citenamefont {Tejada},
  \citenamefont {Chudnovsky}, \citenamefont {del Barco}, \citenamefont
  {Hernandez},\ and\ \citenamefont {Spiller}}]{Tejada2001}%
  \BibitemOpen
  \bibfield  {author} {\bibinfo {author} {\bibfnamefont {J.}~\bibnamefont
  {Tejada}}, \bibinfo {author} {\bibfnamefont {E.~M.}\ \bibnamefont
  {Chudnovsky}}, \bibinfo {author} {\bibfnamefont {E.}~\bibnamefont {del
  Barco}}, \bibinfo {author} {\bibfnamefont {J.~M.}\ \bibnamefont {Hernandez}},
  \ and\ \bibinfo {author} {\bibfnamefont {T.~P.}\ \bibnamefont {Spiller}},\
  }\href {http://stacks.iop.org/0957-4484/12/i=2/a=323} {\bibfield  {journal}
  {\bibinfo  {journal} {Nanotechnology}\ }\textbf {\bibinfo {volume} {12}},\
  \bibinfo {pages} {181} (\bibinfo {year} {2001})}\BibitemShut {NoStop}%
\bibitem [{\citenamefont {Timm}\ and\ \citenamefont {Elste}(2006)}]{Timm2006}%
  \BibitemOpen
  \bibfield  {author} {\bibinfo {author} {\bibfnamefont {C.}~\bibnamefont
  {Timm}}\ and\ \bibinfo {author} {\bibfnamefont {F.}~\bibnamefont {Elste}},\
  }\href {\doibase 10.1103/PhysRevB.73.235304} {\bibfield  {journal} {\bibinfo
  {journal} {Phys. Rev. B}\ }\textbf {\bibinfo {volume} {73}},\ \bibinfo
  {pages} {235304} (\bibinfo {year} {2006})}\BibitemShut {NoStop}%
\bibitem [{\citenamefont {Leuenberger}\ and\ \citenamefont
  {Mucciolo}(2006)}]{Leuenberger2006}%
  \BibitemOpen
  \bibfield  {author} {\bibinfo {author} {\bibfnamefont {M.~N.}\ \bibnamefont
  {Leuenberger}}\ and\ \bibinfo {author} {\bibfnamefont {E.~R.}\ \bibnamefont
  {Mucciolo}},\ }\href {\doibase 10.1103/PhysRevLett.97.126601} {\bibfield
  {journal} {\bibinfo  {journal} {Phys. Rev. Lett.}\ }\textbf {\bibinfo
  {volume} {97}},\ \bibinfo {pages} {126601} (\bibinfo {year}
  {2006})}\BibitemShut {NoStop}%
\bibitem [{\citenamefont {Loss}\ \emph {et~al.}(1992)\citenamefont {Loss},
  \citenamefont {DiVincenzo},\ and\ \citenamefont {Grinstein}}]{Loss1992}%
  \BibitemOpen
  \bibfield  {author} {\bibinfo {author} {\bibfnamefont {D.}~\bibnamefont
  {Loss}}, \bibinfo {author} {\bibfnamefont {D.~P.}\ \bibnamefont
  {DiVincenzo}}, \ and\ \bibinfo {author} {\bibfnamefont {G.}~\bibnamefont
  {Grinstein}},\ }\href {\doibase 10.1103/PhysRevLett.69.3232} {\bibfield
  {journal} {\bibinfo  {journal} {Phys. Rev. Lett.}\ }\textbf {\bibinfo
  {volume} {69}},\ \bibinfo {pages} {3232} (\bibinfo {year}
  {1992})}\BibitemShut {NoStop}%
\bibitem [{\citenamefont {von Delft}\ and\ \citenamefont
  {Henley}(1992)}]{Delft1992}%
  \BibitemOpen
  \bibfield  {author} {\bibinfo {author} {\bibfnamefont {J.}~\bibnamefont {von
  Delft}}\ and\ \bibinfo {author} {\bibfnamefont {C.~L.}\ \bibnamefont
  {Henley}},\ }\href {\doibase 10.1103/PhysRevLett.69.3236} {\bibfield
  {journal} {\bibinfo  {journal} {Phys. Rev. Lett.}\ }\textbf {\bibinfo
  {volume} {69}},\ \bibinfo {pages} {3236} (\bibinfo {year}
  {1992})}\BibitemShut {NoStop}%
\bibitem [{\citenamefont {del Barco}\ \emph {et~al.}(2003)\citenamefont {del
  Barco}, \citenamefont {Kent}, \citenamefont {Rumberger}, \citenamefont
  {Hendrickson},\ and\ \citenamefont {Christou}}]{delBarco2003}%
  \BibitemOpen
  \bibfield  {author} {\bibinfo {author} {\bibfnamefont {E.}~\bibnamefont {del
  Barco}}, \bibinfo {author} {\bibfnamefont {A.~D.}\ \bibnamefont {Kent}},
  \bibinfo {author} {\bibfnamefont {E.~M.}\ \bibnamefont {Rumberger}}, \bibinfo
  {author} {\bibfnamefont {D.~N.}\ \bibnamefont {Hendrickson}}, \ and\ \bibinfo
  {author} {\bibfnamefont {G.}~\bibnamefont {Christou}},\ }\href {\doibase
  10.1103/PhysRevLett.91.047203} {\bibfield  {journal} {\bibinfo  {journal}
  {Phys. Rev. Lett.}\ }\textbf {\bibinfo {volume} {91}},\ \bibinfo {pages}
  {047203} (\bibinfo {year} {2003})}\BibitemShut {NoStop}%
\bibitem [{\citenamefont {Foss-Feig}\ and\ \citenamefont
  {Friedman}(2009)}]{Foss-Feig2009}%
  \BibitemOpen
  \bibfield  {author} {\bibinfo {author} {\bibfnamefont {M.~S.}\ \bibnamefont
  {Foss-Feig}}\ and\ \bibinfo {author} {\bibfnamefont {J.~R.}\ \bibnamefont
  {Friedman}},\ }\href {http://stacks.iop.org/0295-5075/86/i=2/a=27002}
  {\bibfield  {journal} {\bibinfo  {journal} {Europhys. Lett.}\ }\textbf
  {\bibinfo {volume} {86}},\ \bibinfo {pages} {27002} (\bibinfo {year}
  {2009})}\BibitemShut {NoStop}%
\bibitem [{\citenamefont {Atkinson}\ \emph {et~al.}(2014)\citenamefont
  {Atkinson}, \citenamefont {Inglis}, \citenamefont {del Barco},\ and\
  \citenamefont {Brechin}}]{Atkinson2014}%
  \BibitemOpen
  \bibfield  {author} {\bibinfo {author} {\bibfnamefont {J.~H.}\ \bibnamefont
  {Atkinson}}, \bibinfo {author} {\bibfnamefont {R.}~\bibnamefont {Inglis}},
  \bibinfo {author} {\bibfnamefont {E.}~\bibnamefont {del Barco}}, \ and\
  \bibinfo {author} {\bibfnamefont {E.~K.}\ \bibnamefont {Brechin}},\ }\href
  {\doibase 10.1103/PhysRevLett.113.087201} {\bibfield  {journal} {\bibinfo
  {journal} {Phys. Rev. Lett.}\ }\textbf {\bibinfo {volume} {113}},\ \bibinfo
  {pages} {087201} (\bibinfo {year} {2014})}\BibitemShut {NoStop}%
\bibitem [{\citenamefont {Ishikawa}\ \emph {et~al.}(2003)\citenamefont
  {Ishikawa}, \citenamefont {Sugita}, \citenamefont {Ishikawa}, \citenamefont
  {Koshihara},\ and\ \citenamefont {Kaizu}}]{Ishikawa2003}%
  \BibitemOpen
  \bibfield  {author} {\bibinfo {author} {\bibfnamefont {N.}~\bibnamefont
  {Ishikawa}}, \bibinfo {author} {\bibfnamefont {M.}~\bibnamefont {Sugita}},
  \bibinfo {author} {\bibfnamefont {T.}~\bibnamefont {Ishikawa}}, \bibinfo
  {author} {\bibfnamefont {S.}~\bibnamefont {Koshihara}}, \ and\ \bibinfo
  {author} {\bibfnamefont {Y.}~\bibnamefont {Kaizu}},\ }\href {\doibase
  10.1021/ja029629n} {\bibfield  {journal} {\bibinfo  {journal} {J. Am. Chem.
  Soc.}\ }\textbf {\bibinfo {volume} {125}},\ \bibinfo {pages} {8694} (\bibinfo
  {year} {2003})}\BibitemShut {NoStop}%
\bibitem [{\citenamefont {Alam}\ \emph {et~al.}(2006)\citenamefont {Alam},
  \citenamefont {Dremov}, \citenamefont {M\"uller}, \citenamefont {Postnikov},
  \citenamefont {Mal}, \citenamefont {Hussain},\ and\ \citenamefont
  {Kortz}}]{Alam2006}%
  \BibitemOpen
  \bibfield  {author} {\bibinfo {author} {\bibfnamefont {M.~S.}\ \bibnamefont
  {Alam}}, \bibinfo {author} {\bibfnamefont {V.}~\bibnamefont {Dremov}},
  \bibinfo {author} {\bibfnamefont {P.}~\bibnamefont {M\"uller}}, \bibinfo
  {author} {\bibfnamefont {A.~V.}\ \bibnamefont {Postnikov}}, \bibinfo {author}
  {\bibfnamefont {S.~S.}\ \bibnamefont {Mal}}, \bibinfo {author} {\bibfnamefont
  {F.}~\bibnamefont {Hussain}}, \ and\ \bibinfo {author} {\bibfnamefont
  {U.}~\bibnamefont {Kortz}},\ }\href {\doibase 10.1021/ic051586z} {\bibfield
  {journal} {\bibinfo  {journal} {Inorg. Chem.}\ }\textbf {\bibinfo {volume}
  {45}},\ \bibinfo {pages} {2866} (\bibinfo {year} {2006})},\ \bibinfo {note}
  {pMID: 16562942},\ \Eprint
  {http://arxiv.org/abs/http://dx.doi.org/10.1021/ic051586z}
  {http://dx.doi.org/10.1021/ic051586z} \BibitemShut {NoStop}%
\bibitem [{\citenamefont {AlDamen}\ \emph {et~al.}(2008)\citenamefont
  {AlDamen}, \citenamefont {Clemente-Juan}, \citenamefont {Coronado},
  \citenamefont {Mart$\rm\acute{\imath}$-Gastaldo},\ and\ \citenamefont
  {Gaita-Ari$\rm\tilde{n}$o}}]{AlDamen2008}%
  \BibitemOpen
  \bibfield  {author} {\bibinfo {author} {\bibfnamefont {M.~A.}\ \bibnamefont
  {AlDamen}}, \bibinfo {author} {\bibfnamefont {J.~M.}\ \bibnamefont
  {Clemente-Juan}}, \bibinfo {author} {\bibfnamefont {E.}~\bibnamefont
  {Coronado}}, \bibinfo {author} {\bibfnamefont {C.}~\bibnamefont
  {Mart$\rm\acute{\imath}$-Gastaldo}}, \ and\ \bibinfo {author} {\bibfnamefont
  {A.}~\bibnamefont {Gaita-Ari$\rm\tilde{n}$o}},\ }\href {\doibase
  10.1021/ja801659m} {\bibfield  {journal} {\bibinfo  {journal} {J. Am. Chem.
  Soc.}\ }\textbf {\bibinfo {volume} {130}},\ \bibinfo {pages} {8874} (\bibinfo
  {year} {2008})},\ \bibinfo {note} {pMID: 18558687},\ \Eprint
  {http://arxiv.org/abs/http://dx.doi.org/10.1021/ja801659m}
  {http://dx.doi.org/10.1021/ja801659m} \BibitemShut {NoStop}%
\bibitem [{\citenamefont {Zadrozny}\ \emph {et~al.}(2013)\citenamefont
  {Zadrozny}, \citenamefont {Xiao}, \citenamefont {Atanasov}, \citenamefont
  {Long}, \citenamefont {Grandjean}, \citenamefont {Neese},\ and\ \citenamefont
  {Long}}]{Zadrozny2013}%
  \BibitemOpen
  \bibfield  {author} {\bibinfo {author} {\bibfnamefont {J.~M.}\ \bibnamefont
  {Zadrozny}}, \bibinfo {author} {\bibfnamefont {D.~J.}\ \bibnamefont {Xiao}},
  \bibinfo {author} {\bibfnamefont {M.}~\bibnamefont {Atanasov}}, \bibinfo
  {author} {\bibfnamefont {G.~J.}\ \bibnamefont {Long}}, \bibinfo {author}
  {\bibfnamefont {F.}~\bibnamefont {Grandjean}}, \bibinfo {author}
  {\bibfnamefont {F.}~\bibnamefont {Neese}}, \ and\ \bibinfo {author}
  {\bibfnamefont {J.~R.}\ \bibnamefont {Long}},\ }\href {\doibase
  10.1038/nchem.1630} {\bibfield  {journal} {\bibinfo  {journal} {Nat. Chem.}\
  }\textbf {\bibinfo {volume} {5}},\ \bibinfo {pages} {577} (\bibinfo {year}
  {2013})}\BibitemShut {NoStop}%
\bibitem [{\citenamefont {Kazin}\ \emph {et~al.}(2014)\citenamefont {Kazin},
  \citenamefont {Zykin}, \citenamefont {Schnelle}, \citenamefont {Felser},\
  and\ \citenamefont {Jansen}}]{Kazin2014}%
  \BibitemOpen
  \bibfield  {author} {\bibinfo {author} {\bibfnamefont {P.~E.}\ \bibnamefont
  {Kazin}}, \bibinfo {author} {\bibfnamefont {M.~A.}\ \bibnamefont {Zykin}},
  \bibinfo {author} {\bibfnamefont {W.}~\bibnamefont {Schnelle}}, \bibinfo
  {author} {\bibfnamefont {C.}~\bibnamefont {Felser}}, \ and\ \bibinfo {author}
  {\bibfnamefont {M.}~\bibnamefont {Jansen}},\ }\href {\doibase
  10.1039/C4CC03966A} {\bibfield  {journal} {\bibinfo  {journal} {Chem.
  Commun.}\ }\textbf {\bibinfo {volume} {50}},\ \bibinfo {pages} {9325}
  (\bibinfo {year} {2014})}\BibitemShut {NoStop}%
\bibitem [{\citenamefont {Liu}\ \emph {et~al.}(2016)\citenamefont {Liu},
  \citenamefont {Chen}, \citenamefont {Liu}, \citenamefont {Vieru},
  \citenamefont {Ungur}, \citenamefont {Jia}, \citenamefont {Chibotaru},
  \citenamefont {Lan}, \citenamefont {Wernsdorfer}, \citenamefont {Gao},
  \citenamefont {Chen},\ and\ \citenamefont {Tong}}]{Jiang2016}%
  \BibitemOpen
  \bibfield  {author} {\bibinfo {author} {\bibfnamefont {J.}~\bibnamefont
  {Liu}}, \bibinfo {author} {\bibfnamefont {Y.-C.}\ \bibnamefont {Chen}},
  \bibinfo {author} {\bibfnamefont {J.-L.}\ \bibnamefont {Liu}}, \bibinfo
  {author} {\bibfnamefont {V.}~\bibnamefont {Vieru}}, \bibinfo {author}
  {\bibfnamefont {L.}~\bibnamefont {Ungur}}, \bibinfo {author} {\bibfnamefont
  {J.-H.}\ \bibnamefont {Jia}}, \bibinfo {author} {\bibfnamefont {L.~F.}\
  \bibnamefont {Chibotaru}}, \bibinfo {author} {\bibfnamefont {Y.}~\bibnamefont
  {Lan}}, \bibinfo {author} {\bibfnamefont {W.}~\bibnamefont {Wernsdorfer}},
  \bibinfo {author} {\bibfnamefont {S.}~\bibnamefont {Gao}}, \bibinfo {author}
  {\bibfnamefont {X.-M.}\ \bibnamefont {Chen}}, \ and\ \bibinfo {author}
  {\bibfnamefont {M.-L.}\ \bibnamefont {Tong}},\ }\href {\doibase
  10.1021/jacs.6b02638} {\bibfield  {journal} {\bibinfo  {journal} {J. Am.
  Chem. Soc.}\ }\textbf {\bibinfo {volume} {138}},\ \bibinfo {pages} {5441}
  (\bibinfo {year} {2016})},\ \bibinfo {note} {pMID: 27054904},\ \Eprint
  {http://arxiv.org/abs/http://dx.doi.org/10.1021/jacs.6b02638}
  {http://dx.doi.org/10.1021/jacs.6b02638} \BibitemShut {NoStop}%
\bibitem [{\citenamefont {Giraud}\ \emph {et~al.}(2001)\citenamefont {Giraud},
  \citenamefont {Wernsdorfer}, \citenamefont {Tkachuk}, \citenamefont
  {Mailly},\ and\ \citenamefont {Barbara}}]{Giraud2001}%
  \BibitemOpen
  \bibfield  {author} {\bibinfo {author} {\bibfnamefont {R.}~\bibnamefont
  {Giraud}}, \bibinfo {author} {\bibfnamefont {W.}~\bibnamefont {Wernsdorfer}},
  \bibinfo {author} {\bibfnamefont {A.~M.}\ \bibnamefont {Tkachuk}}, \bibinfo
  {author} {\bibfnamefont {D.}~\bibnamefont {Mailly}}, \ and\ \bibinfo {author}
  {\bibfnamefont {B.}~\bibnamefont {Barbara}},\ }\href {\doibase
  10.1103/PhysRevLett.87.057203} {\bibfield  {journal} {\bibinfo  {journal}
  {Phys. Rev. Lett.}\ }\textbf {\bibinfo {volume} {87}},\ \bibinfo {pages}
  {057203} (\bibinfo {year} {2001})}\BibitemShut {NoStop}%
\bibitem [{\citenamefont {Barbara}\ \emph {et~al.}(2004)\citenamefont
  {Barbara}, \citenamefont {Giraud}, \citenamefont {Wernsdorfer}, \citenamefont
  {Mailly}, \citenamefont {Lejay}, \citenamefont {Tkachuk},\ and\ \citenamefont
  {Suzuki}}]{Barbara2004}%
  \BibitemOpen
  \bibfield  {author} {\bibinfo {author} {\bibfnamefont {B.}~\bibnamefont
  {Barbara}}, \bibinfo {author} {\bibfnamefont {R.}~\bibnamefont {Giraud}},
  \bibinfo {author} {\bibfnamefont {W.}~\bibnamefont {Wernsdorfer}}, \bibinfo
  {author} {\bibfnamefont {D.}~\bibnamefont {Mailly}}, \bibinfo {author}
  {\bibfnamefont {P.}~\bibnamefont {Lejay}}, \bibinfo {author} {\bibfnamefont
  {A.}~\bibnamefont {Tkachuk}}, \ and\ \bibinfo {author} {\bibfnamefont
  {H.}~\bibnamefont {Suzuki}},\ }\href {\doibase
  https://doi.org/10.1016/j.jmmm.2003.12.654} {\bibfield  {journal} {\bibinfo
  {journal} {Journal of Magnetism and Magnetic Materials}\ }\textbf {\bibinfo
  {volume} {272-276}},\ \bibinfo {pages} {1024 } (\bibinfo {year}
  {2004})}\BibitemShut {NoStop}%
\bibitem [{\citenamefont {Giraud}\ \emph {et~al.}(2003)\citenamefont {Giraud},
  \citenamefont {Tkachuk},\ and\ \citenamefont {Barbara}}]{Giraud2003}%
  \BibitemOpen
  \bibfield  {author} {\bibinfo {author} {\bibfnamefont {R.}~\bibnamefont
  {Giraud}}, \bibinfo {author} {\bibfnamefont {A.~M.}\ \bibnamefont {Tkachuk}},
  \ and\ \bibinfo {author} {\bibfnamefont {B.}~\bibnamefont {Barbara}},\
  }\href@noop {} {\bibfield  {journal} {\bibinfo  {journal} {Phys. Rev. Lett.}\
  }\textbf {\bibinfo {volume} {91}},\ \bibinfo {pages} {257204} (\bibinfo
  {year} {2003})}\BibitemShut {NoStop}%
\bibitem [{\citenamefont {Bertaina}\ \emph {et~al.}(2006)\citenamefont
  {Bertaina}, \citenamefont {Barbara}, \citenamefont {Giraud}, \citenamefont
  {Malkin}, \citenamefont {Vanuynin}, \citenamefont {Pominov}, \citenamefont
  {Stolov},\ and\ \citenamefont {Tkachuk}}]{Bertaina2006}%
  \BibitemOpen
  \bibfield  {author} {\bibinfo {author} {\bibfnamefont {S.}~\bibnamefont
  {Bertaina}}, \bibinfo {author} {\bibfnamefont {B.}~\bibnamefont {Barbara}},
  \bibinfo {author} {\bibfnamefont {R.}~\bibnamefont {Giraud}}, \bibinfo
  {author} {\bibfnamefont {B.~Z.}\ \bibnamefont {Malkin}}, \bibinfo {author}
  {\bibfnamefont {M.~V.}\ \bibnamefont {Vanuynin}}, \bibinfo {author}
  {\bibfnamefont {A.~I.}\ \bibnamefont {Pominov}}, \bibinfo {author}
  {\bibfnamefont {A.~L.}\ \bibnamefont {Stolov}}, \ and\ \bibinfo {author}
  {\bibfnamefont {A.~M.}\ \bibnamefont {Tkachuk}},\ }\href {\doibase
  10.1103/PhysRevB.74.184421} {\bibfield  {journal} {\bibinfo  {journal} {Phys.
  Rev. B}\ }\textbf {\bibinfo {volume} {74}},\ \bibinfo {pages} {184421}
  (\bibinfo {year} {2006})}\BibitemShut {NoStop}%
\bibitem [{\citenamefont {Jesche}\ \emph {et~al.}(2014)\citenamefont {Jesche},
  \citenamefont {McCallum}, \citenamefont {Thimmaiah}, \citenamefont {Jacobs},
  \citenamefont {Taufour}, \citenamefont {Kreyssig}, \citenamefont {Houk},
  \citenamefont {Bud'ko},\ and\ \citenamefont {Canfield}}]{Jesche2014b}%
  \BibitemOpen
  \bibfield  {author} {\bibinfo {author} {\bibfnamefont {A.}~\bibnamefont
  {Jesche}}, \bibinfo {author} {\bibfnamefont {R.~W.}\ \bibnamefont
  {McCallum}}, \bibinfo {author} {\bibfnamefont {S.}~\bibnamefont {Thimmaiah}},
  \bibinfo {author} {\bibfnamefont {J.~L.}\ \bibnamefont {Jacobs}}, \bibinfo
  {author} {\bibfnamefont {V.}~\bibnamefont {Taufour}}, \bibinfo {author}
  {\bibfnamefont {A.}~\bibnamefont {Kreyssig}}, \bibinfo {author}
  {\bibfnamefont {R.~S.}\ \bibnamefont {Houk}}, \bibinfo {author}
  {\bibfnamefont {S.~L.}\ \bibnamefont {Bud'ko}}, \ and\ \bibinfo {author}
  {\bibfnamefont {P.~C.}\ \bibnamefont {Canfield}},\ }\href
  {http://dx.doi.org/10.1038/ncomms4333} {\bibfield  {journal} {\bibinfo
  {journal} {Nat. Commun.}\ }\textbf {\bibinfo {volume} {5:3333}} (\bibinfo
  {year} {2014})},\ \bibinfo {note} {doi: 10.1038/ncomms4333}\BibitemShut
  {NoStop}%
\bibitem [{\citenamefont {Fix}\ \emph {et~al.}(2017)\citenamefont {Fix},
  \citenamefont {Jesche}, \citenamefont {Jantz}, \citenamefont {Br\"auninger},
  \citenamefont {Klauss}, \citenamefont {Manna}, \citenamefont {Pietsch},
  \citenamefont {H\"oppe},\ and\ \citenamefont {Canfield}}]{Fix2017}%
  \BibitemOpen
  \bibfield  {author} {\bibinfo {author} {\bibfnamefont {M.}~\bibnamefont
  {Fix}}, \bibinfo {author} {\bibfnamefont {A.}~\bibnamefont {Jesche}},
  \bibinfo {author} {\bibfnamefont {S.~G.}\ \bibnamefont {Jantz}}, \bibinfo
  {author} {\bibfnamefont {S.~A.}\ \bibnamefont {Br\"auninger}}, \bibinfo
  {author} {\bibfnamefont {H.-H.}\ \bibnamefont {Klauss}}, \bibinfo {author}
  {\bibfnamefont {R.~S.}\ \bibnamefont {Manna}}, \bibinfo {author}
  {\bibfnamefont {I.~M.}\ \bibnamefont {Pietsch}}, \bibinfo {author}
  {\bibfnamefont {H.~A.}\ \bibnamefont {H\"oppe}}, \ and\ \bibinfo {author}
  {\bibfnamefont {P.~C.}\ \bibnamefont {Canfield}},\ }\href
  {http://arxiv.org/abs/1712.07953} {\bibfield  {journal} {\bibinfo  {journal}
  {arXiv:1712.07953 [cond-mat]}\ } (\bibinfo {year} {2017})}\BibitemShut
  {NoStop}%
\bibitem [{\citenamefont {Jesche}\ \emph {et~al.}(2015)\citenamefont {Jesche},
  \citenamefont {Ke}, \citenamefont {Jacobs}, \citenamefont {Harmon},
  \citenamefont {Houk},\ and\ \citenamefont {Canfield}}]{Jesche2015}%
  \BibitemOpen
  \bibfield  {author} {\bibinfo {author} {\bibfnamefont {A.}~\bibnamefont
  {Jesche}}, \bibinfo {author} {\bibfnamefont {L.}~\bibnamefont {Ke}}, \bibinfo
  {author} {\bibfnamefont {J.~L.}\ \bibnamefont {Jacobs}}, \bibinfo {author}
  {\bibfnamefont {B.}~\bibnamefont {Harmon}}, \bibinfo {author} {\bibfnamefont
  {R.~S.}\ \bibnamefont {Houk}}, \ and\ \bibinfo {author} {\bibfnamefont
  {P.~C.}\ \bibnamefont {Canfield}},\ }\href {\doibase
  10.1103/PhysRevB.91.180403} {\bibfield  {journal} {\bibinfo  {journal} {Phys.
  Rev. B}\ }\textbf {\bibinfo {volume} {91}},\ \bibinfo {pages} {180403}
  (\bibinfo {year} {2015})}\BibitemShut {NoStop}%
\bibitem [{\citenamefont {Klatyk}\ \emph {et~al.}(2002)\citenamefont {Klatyk},
  \citenamefont {Schnelle}, \citenamefont {Wagner}, \citenamefont {Niewa},
  \citenamefont {Nov\'ak}, \citenamefont {Kniep}, \citenamefont {Waldeck},
  \citenamefont {Ksenofontov},\ and\ \citenamefont {G\"utlich}}]{Klatyk2002}%
  \BibitemOpen
  \bibfield  {author} {\bibinfo {author} {\bibfnamefont {J.}~\bibnamefont
  {Klatyk}}, \bibinfo {author} {\bibfnamefont {W.}~\bibnamefont {Schnelle}},
  \bibinfo {author} {\bibfnamefont {F.~R.}\ \bibnamefont {Wagner}}, \bibinfo
  {author} {\bibfnamefont {R.}~\bibnamefont {Niewa}}, \bibinfo {author}
  {\bibfnamefont {P.}~\bibnamefont {Nov\'ak}}, \bibinfo {author} {\bibfnamefont
  {R.}~\bibnamefont {Kniep}}, \bibinfo {author} {\bibfnamefont
  {M.}~\bibnamefont {Waldeck}}, \bibinfo {author} {\bibfnamefont
  {V.}~\bibnamefont {Ksenofontov}}, \ and\ \bibinfo {author} {\bibfnamefont
  {P.}~\bibnamefont {G\"utlich}},\ }\href@noop {} {\bibfield  {journal}
  {\bibinfo  {journal} {Phys. Rev. Lett.}\ }\textbf {\bibinfo {volume} {88}},\
  \bibinfo {pages} {207202} (\bibinfo {year} {2002})}\BibitemShut {NoStop}%
\bibitem [{\citenamefont {Nov\'ak}\ and\ \citenamefont
  {Wagner}(2002)}]{Novak2002}%
  \BibitemOpen
  \bibfield  {author} {\bibinfo {author} {\bibfnamefont {P.}~\bibnamefont
  {Nov\'ak}}\ and\ \bibinfo {author} {\bibfnamefont {F.~R.}\ \bibnamefont
  {Wagner}},\ }\href {\doibase 10.1103/PhysRevB.66.184434} {\bibfield
  {journal} {\bibinfo  {journal} {Phys. Rev. B}\ }\textbf {\bibinfo {volume}
  {66}},\ \bibinfo {pages} {184434} (\bibinfo {year} {2002})}\BibitemShut
  {NoStop}%
\bibitem [{\citenamefont {Antropov}\ and\ \citenamefont
  {Antonov}(2014)}]{Antropov2014}%
  \BibitemOpen
  \bibfield  {author} {\bibinfo {author} {\bibfnamefont {V.~P.}\ \bibnamefont
  {Antropov}}\ and\ \bibinfo {author} {\bibfnamefont {V.~N.}\ \bibnamefont
  {Antonov}},\ }\href {\doibase 10.1103/PhysRevB.90.094406} {\bibfield
  {journal} {\bibinfo  {journal} {Phys. Rev. B}\ }\textbf {\bibinfo {volume}
  {90}},\ \bibinfo {pages} {094406} (\bibinfo {year} {2014})}\BibitemShut
  {NoStop}%
\bibitem [{\citenamefont {Ke}\ and\ \citenamefont {van
  Schilfgaarde}(2015)}]{Ke2015}%
  \BibitemOpen
  \bibfield  {author} {\bibinfo {author} {\bibfnamefont {L.}~\bibnamefont
  {Ke}}\ and\ \bibinfo {author} {\bibfnamefont {M.}~\bibnamefont {van
  Schilfgaarde}},\ }\href {\doibase 10.1103/PhysRevB.92.014423} {\bibfield
  {journal} {\bibinfo  {journal} {Phys. Rev. B}\ }\textbf {\bibinfo {volume}
  {92}},\ \bibinfo {pages} {014423} (\bibinfo {year} {2015})}\BibitemShut
  {NoStop}%
\bibitem [{\citenamefont {Xu}\ \emph {et~al.}(2017)\citenamefont {Xu},
  \citenamefont {Zangeneh}, \citenamefont {Yadav}, \citenamefont {Avdoshenko},
  \citenamefont {van~den Brink}, \citenamefont {Jesche},\ and\ \citenamefont
  {Hozoi}}]{Xu2017}%
  \BibitemOpen
  \bibfield  {author} {\bibinfo {author} {\bibfnamefont {L.}~\bibnamefont
  {Xu}}, \bibinfo {author} {\bibfnamefont {Z.}~\bibnamefont {Zangeneh}},
  \bibinfo {author} {\bibfnamefont {R.}~\bibnamefont {Yadav}}, \bibinfo
  {author} {\bibfnamefont {S.}~\bibnamefont {Avdoshenko}}, \bibinfo {author}
  {\bibfnamefont {J.}~\bibnamefont {van~den Brink}}, \bibinfo {author}
  {\bibfnamefont {A.}~\bibnamefont {Jesche}}, \ and\ \bibinfo {author}
  {\bibfnamefont {L.}~\bibnamefont {Hozoi}},\ }\href {\doibase
  10.1039/C7NR03041J} {\bibfield  {journal} {\bibinfo  {journal} {Nanoscale}\
  }\textbf {\bibinfo {volume} {9}},\ \bibinfo {pages} {10596} (\bibinfo {year}
  {2017})}\BibitemShut {NoStop}%
\bibitem [{\citenamefont {Jesche}\ and\ \citenamefont
  {Canfield}(2014)}]{Jesche2014c}%
  \BibitemOpen
  \bibfield  {author} {\bibinfo {author} {\bibfnamefont {A.}~\bibnamefont
  {Jesche}}\ and\ \bibinfo {author} {\bibfnamefont {P.~C.}\ \bibnamefont
  {Canfield}},\ }\href {\doibase 10.1080/14786435.2014.913114} {\bibfield
  {journal} {\bibinfo  {journal} {Philos. Mag.}\ }\textbf {\bibinfo {volume}
  {94}},\ \bibinfo {pages} {2372} (\bibinfo {year} {2014})}\BibitemShut
  {NoStop}%
\bibitem [{SI()}]{SI}%
  \BibitemOpen
  \href@noop {} {}\bibinfo {note} {See Supplemental Material at [URL will be
  added] for further details on transverse measurements, the remnant field of
  the MPMS3 magnet, magnetic ac susceptibility data and details on the analysis
  of the time-dependent magnetization.}\BibitemShut {Stop}%
\bibitem [{\citenamefont {Zener}(1932)}]{Zener1932}%
  \BibitemOpen
  \bibfield  {author} {\bibinfo {author} {\bibfnamefont {C.}~\bibnamefont
  {Zener}},\ }\href {\doibase 10.1098/rspa.1932.0165} {\bibfield  {journal}
  {\bibinfo  {journal} {Proc. R. Soc. A}\ }\textbf {\bibinfo {volume} {137}},\
  \bibinfo {pages} {696} (\bibinfo {year} {1932})}\BibitemShut {NoStop}%
\bibitem [{\citenamefont {Sangregorio}\ \emph {et~al.}(1997)\citenamefont
  {Sangregorio}, \citenamefont {Ohm}, \citenamefont {Paulsen}, \citenamefont
  {Sessoli},\ and\ \citenamefont {Gatteschi}}]{Sangregorio1997}%
  \BibitemOpen
  \bibfield  {author} {\bibinfo {author} {\bibfnamefont {C.}~\bibnamefont
  {Sangregorio}}, \bibinfo {author} {\bibfnamefont {T.}~\bibnamefont {Ohm}},
  \bibinfo {author} {\bibfnamefont {C.}~\bibnamefont {Paulsen}}, \bibinfo
  {author} {\bibfnamefont {R.}~\bibnamefont {Sessoli}}, \ and\ \bibinfo
  {author} {\bibfnamefont {D.}~\bibnamefont {Gatteschi}},\ }\href {\doibase
  10.1103/PhysRevLett.78.4645} {\bibfield  {journal} {\bibinfo  {journal}
  {Phys. Rev. Lett.}\ }\textbf {\bibinfo {volume} {78}},\ \bibinfo {pages}
  {4645} (\bibinfo {year} {1997})}\BibitemShut {NoStop}%
\bibitem [{\citenamefont {Friedman}\ \emph {et~al.}(1998)\citenamefont
  {Friedman}, \citenamefont {Sarachik},\ and\ \citenamefont
  {Ziolo}}]{Friedman1998}%
  \BibitemOpen
  \bibfield  {author} {\bibinfo {author} {\bibfnamefont {J.~R.}\ \bibnamefont
  {Friedman}}, \bibinfo {author} {\bibfnamefont {M.~P.}\ \bibnamefont
  {Sarachik}}, \ and\ \bibinfo {author} {\bibfnamefont {R.}~\bibnamefont
  {Ziolo}},\ }\href {\doibase 10.1103/PhysRevB.58.R14729} {\bibfield  {journal}
  {\bibinfo  {journal} {Phys. Rev. B}\ }\textbf {\bibinfo {volume} {58}},\
  \bibinfo {pages} {R14729} (\bibinfo {year} {1998})}\BibitemShut {NoStop}%
\bibitem [{\citenamefont {Fort}\ \emph {et~al.}(1998)\citenamefont {Fort},
  \citenamefont {Rettori}, \citenamefont {Villain}, \citenamefont {Gatteschi},\
  and\ \citenamefont {Sessoli}}]{Fort1998}%
  \BibitemOpen
  \bibfield  {author} {\bibinfo {author} {\bibfnamefont {A.}~\bibnamefont
  {Fort}}, \bibinfo {author} {\bibfnamefont {A.}~\bibnamefont {Rettori}},
  \bibinfo {author} {\bibfnamefont {J.}~\bibnamefont {Villain}}, \bibinfo
  {author} {\bibfnamefont {D.}~\bibnamefont {Gatteschi}}, \ and\ \bibinfo
  {author} {\bibfnamefont {R.}~\bibnamefont {Sessoli}},\ }\href {\doibase
  10.1103/PhysRevLett.80.612} {\bibfield  {journal} {\bibinfo  {journal} {Phys.
  Rev. Lett.}\ }\textbf {\bibinfo {volume} {80}},\ \bibinfo {pages} {612}
  (\bibinfo {year} {1998})}\BibitemShut {NoStop}%
\bibitem [{\citenamefont {Fominaya}\ \emph {et~al.}(1999)\citenamefont
  {Fominaya}, \citenamefont {Villain}, \citenamefont {Fournier}, \citenamefont
  {Gandit}, \citenamefont {Chaussy}, \citenamefont {Fort},\ and\ \citenamefont
  {Caneschi}}]{Fominaya1999}%
  \BibitemOpen
  \bibfield  {author} {\bibinfo {author} {\bibfnamefont {F.}~\bibnamefont
  {Fominaya}}, \bibinfo {author} {\bibfnamefont {J.}~\bibnamefont {Villain}},
  \bibinfo {author} {\bibfnamefont {T.}~\bibnamefont {Fournier}}, \bibinfo
  {author} {\bibfnamefont {P.}~\bibnamefont {Gandit}}, \bibinfo {author}
  {\bibfnamefont {J.}~\bibnamefont {Chaussy}}, \bibinfo {author} {\bibfnamefont
  {A.}~\bibnamefont {Fort}}, \ and\ \bibinfo {author} {\bibfnamefont
  {A.}~\bibnamefont {Caneschi}},\ }\href {\doibase 10.1103/PhysRevB.59.519}
  {\bibfield  {journal} {\bibinfo  {journal} {Phys. Rev. B}\ }\textbf {\bibinfo
  {volume} {59}},\ \bibinfo {pages} {519} (\bibinfo {year} {1999})}\BibitemShut
  {NoStop}%
\bibitem [{\citenamefont {Freedman}\ \emph {et~al.}(2010)\citenamefont
  {Freedman}, \citenamefont {Harman}, \citenamefont {Harris}, \citenamefont
  {Long}, \citenamefont {Chang},\ and\ \citenamefont {Long}}]{Freedman2010}%
  \BibitemOpen
  \bibfield  {author} {\bibinfo {author} {\bibfnamefont {D.~E.}\ \bibnamefont
  {Freedman}}, \bibinfo {author} {\bibfnamefont {W.~H.}\ \bibnamefont
  {Harman}}, \bibinfo {author} {\bibfnamefont {T.~D.}\ \bibnamefont {Harris}},
  \bibinfo {author} {\bibfnamefont {G.~J.}\ \bibnamefont {Long}}, \bibinfo
  {author} {\bibfnamefont {C.~J.}\ \bibnamefont {Chang}}, \ and\ \bibinfo
  {author} {\bibfnamefont {J.~R.}\ \bibnamefont {Long}},\ }\href {\doibase
  10.1021/ja909560d} {\bibfield  {journal} {\bibinfo  {journal} {J. Am. Chem.
  Soc.}\ }\textbf {\bibinfo {volume} {132}},\ \bibinfo {pages} {1224} (\bibinfo
  {year} {2010})}\BibitemShut {NoStop}%
\bibitem [{\citenamefont {Fataftah}\ \emph {et~al.}(2014)\citenamefont
  {Fataftah}, \citenamefont {Zadrozny}, \citenamefont {Rogers},\ and\
  \citenamefont {Freedman}}]{Fataftah2014}%
  \BibitemOpen
  \bibfield  {author} {\bibinfo {author} {\bibfnamefont {M.~S.}\ \bibnamefont
  {Fataftah}}, \bibinfo {author} {\bibfnamefont {J.~M.}\ \bibnamefont
  {Zadrozny}}, \bibinfo {author} {\bibfnamefont {D.~M.}\ \bibnamefont
  {Rogers}}, \ and\ \bibinfo {author} {\bibfnamefont {D.~E.}\ \bibnamefont
  {Freedman}},\ }\href {\doibase 10.1021/ic501906z} {\bibfield  {journal}
  {\bibinfo  {journal} {Inorganic Chemistry}\ }\textbf {\bibinfo {volume}
  {53}},\ \bibinfo {pages} {10716} (\bibinfo {year} {2014})},\ \bibinfo {note}
  {pMID: 25198379},\ \Eprint
  {http://arxiv.org/abs/http://dx.doi.org/10.1021/ic501906z}
  {http://dx.doi.org/10.1021/ic501906z} \BibitemShut {NoStop}%
\bibitem [{\citenamefont {Orbach}(1961)}]{Orbach1961}%
  \BibitemOpen
  \bibfield  {author} {\bibinfo {author} {\bibfnamefont {R.}~\bibnamefont
  {Orbach}},\ }\href {\doibase 10.1098/rspa.1961.0211} {\bibfield  {journal}
  {\bibinfo  {journal} {Proc. R. Soc. Lond. A}\ }\textbf {\bibinfo {volume}
  {264}},\ \bibinfo {pages} {458} (\bibinfo {year} {1961})}\BibitemShut
  {NoStop}%
\bibitem [{\citenamefont {Carlin}(1986)}]{Carlin1986}%
  \BibitemOpen
  \bibfield  {author} {\bibinfo {author} {\bibfnamefont {R.~L.}\ \bibnamefont
  {Carlin}},\ }\href@noop {} {\emph {\bibinfo {title} {Magnetochemistry}}}\
  (\bibinfo  {publisher} {Springer Verlag, Berlin Heidelberg},\ \bibinfo {year}
  {1986})\BibitemShut {NoStop}%
\bibitem [{\citenamefont {Zhong}\ \emph {et~al.}(2000)\citenamefont {Zhong},
  \citenamefont {Sarachik}, \citenamefont {Yoo},\ and\ \citenamefont
  {Hendrickson}}]{Zhong2000}%
  \BibitemOpen
  \bibfield  {author} {\bibinfo {author} {\bibfnamefont {Y.}~\bibnamefont
  {Zhong}}, \bibinfo {author} {\bibfnamefont {M.~P.}\ \bibnamefont {Sarachik}},
  \bibinfo {author} {\bibfnamefont {J.}~\bibnamefont {Yoo}}, \ and\ \bibinfo
  {author} {\bibfnamefont {D.~N.}\ \bibnamefont {Hendrickson}},\ }\href
  {\doibase 10.1103/PhysRevB.62.R9256} {\bibfield  {journal} {\bibinfo
  {journal} {Phys. Rev. B}\ }\textbf {\bibinfo {volume} {62}},\ \bibinfo
  {pages} {R9256} (\bibinfo {year} {2000})}\BibitemShut {NoStop}%
\bibitem [{\citenamefont {McHugh}\ \emph {et~al.}(2009)\citenamefont {McHugh},
  \citenamefont {Jaafar}, \citenamefont {Sarachik}, \citenamefont {Myasoedov},
  \citenamefont {Shtrikman}, \citenamefont {Zeldov}, \citenamefont {Bagai},\
  and\ \citenamefont {Christou}}]{McHugh2009}%
  \BibitemOpen
  \bibfield  {author} {\bibinfo {author} {\bibfnamefont {S.}~\bibnamefont
  {McHugh}}, \bibinfo {author} {\bibfnamefont {R.}~\bibnamefont {Jaafar}},
  \bibinfo {author} {\bibfnamefont {M.~P.}\ \bibnamefont {Sarachik}}, \bibinfo
  {author} {\bibfnamefont {Y.}~\bibnamefont {Myasoedov}}, \bibinfo {author}
  {\bibfnamefont {H.}~\bibnamefont {Shtrikman}}, \bibinfo {author}
  {\bibfnamefont {E.}~\bibnamefont {Zeldov}}, \bibinfo {author} {\bibfnamefont
  {R.}~\bibnamefont {Bagai}}, \ and\ \bibinfo {author} {\bibfnamefont
  {G.}~\bibnamefont {Christou}},\ }\href {\doibase 10.1103/PhysRevB.79.052404}
  {\bibfield  {journal} {\bibinfo  {journal} {Phys. Rev. B}\ }\textbf {\bibinfo
  {volume} {79}},\ \bibinfo {pages} {052404} (\bibinfo {year}
  {2009})}\BibitemShut {NoStop}%
\bibitem [{\citenamefont {Chudnovsky}\ and\ \citenamefont
  {Garanin}(2001)}]{Chudnovsky2001}%
  \BibitemOpen
  \bibfield  {author} {\bibinfo {author} {\bibfnamefont {E.~M.}\ \bibnamefont
  {Chudnovsky}}\ and\ \bibinfo {author} {\bibfnamefont {D.~A.}\ \bibnamefont
  {Garanin}},\ }\href {\doibase 10.1103/PhysRevLett.87.187203} {\bibfield
  {journal} {\bibinfo  {journal} {Phys. Rev. Lett.}\ }\textbf {\bibinfo
  {volume} {87}},\ \bibinfo {pages} {187203} (\bibinfo {year}
  {2001})}\BibitemShut {NoStop}%
\bibitem [{\citenamefont {Bertaina}\ \emph {et~al.}(2009)\citenamefont
  {Bertaina}, \citenamefont {Shim}, \citenamefont {Gambarelli}, \citenamefont
  {Malkin},\ and\ \citenamefont {Barbara}}]{Bertaina2009}%
  \BibitemOpen
  \bibfield  {author} {\bibinfo {author} {\bibfnamefont {S.}~\bibnamefont
  {Bertaina}}, \bibinfo {author} {\bibfnamefont {J.~H.}\ \bibnamefont {Shim}},
  \bibinfo {author} {\bibfnamefont {S.}~\bibnamefont {Gambarelli}}, \bibinfo
  {author} {\bibfnamefont {B.~Z.}\ \bibnamefont {Malkin}}, \ and\ \bibinfo
  {author} {\bibfnamefont {B.}~\bibnamefont {Barbara}},\ }\href {\doibase
  10.1103/PhysRevLett.103.226402} {\bibfield  {journal} {\bibinfo  {journal}
  {Phys. Rev. Lett.}\ }\textbf {\bibinfo {volume} {103}},\ \bibinfo {pages}
  {226402} (\bibinfo {year} {2009})}\BibitemShut {NoStop}%
\bibitem [{\citenamefont {Rakhmatullin}\ \emph {et~al.}(2009)\citenamefont
  {Rakhmatullin}, \citenamefont {Kurkin}, \citenamefont {Mamin}, \citenamefont
  {Orlinskii}, \citenamefont {Gafurov}, \citenamefont {Baibekov}, \citenamefont
  {Malkin}, \citenamefont {Gambarelli}, \citenamefont {Bertaina},\ and\
  \citenamefont {Barbara}}]{Rakhmatullin2009}%
  \BibitemOpen
  \bibfield  {author} {\bibinfo {author} {\bibfnamefont {R.~M.}\ \bibnamefont
  {Rakhmatullin}}, \bibinfo {author} {\bibfnamefont {I.~N.}\ \bibnamefont
  {Kurkin}}, \bibinfo {author} {\bibfnamefont {G.~V.}\ \bibnamefont {Mamin}},
  \bibinfo {author} {\bibfnamefont {S.~B.}\ \bibnamefont {Orlinskii}}, \bibinfo
  {author} {\bibfnamefont {M.~R.}\ \bibnamefont {Gafurov}}, \bibinfo {author}
  {\bibfnamefont {E.~I.}\ \bibnamefont {Baibekov}}, \bibinfo {author}
  {\bibfnamefont {B.~Z.}\ \bibnamefont {Malkin}}, \bibinfo {author}
  {\bibfnamefont {S.}~\bibnamefont {Gambarelli}}, \bibinfo {author}
  {\bibfnamefont {S.}~\bibnamefont {Bertaina}}, \ and\ \bibinfo {author}
  {\bibfnamefont {B.}~\bibnamefont {Barbara}},\ }\href {\doibase
  10.1103/PhysRevB.79.172408} {\bibfield  {journal} {\bibinfo  {journal}
  {Physical Review B}\ }\textbf {\bibinfo {volume} {79}},\ \bibinfo {pages}
  {172408} (\bibinfo {year} {2009})}\BibitemShut {NoStop}%
\bibitem [{\citenamefont {Bertaina}\ \emph {et~al.}(2007)\citenamefont
  {Bertaina}, \citenamefont {Gambarelli}, \citenamefont {Tkachuk},
  \citenamefont {Kurkin}, \citenamefont {Malkin}, \citenamefont {Stepanov},\
  and\ \citenamefont {Barbara}}]{Bertaina2007}%
  \BibitemOpen
  \bibfield  {author} {\bibinfo {author} {\bibfnamefont {S.}~\bibnamefont
  {Bertaina}}, \bibinfo {author} {\bibfnamefont {S.}~\bibnamefont
  {Gambarelli}}, \bibinfo {author} {\bibfnamefont {A.}~\bibnamefont {Tkachuk}},
  \bibinfo {author} {\bibfnamefont {I.~N.}\ \bibnamefont {Kurkin}}, \bibinfo
  {author} {\bibfnamefont {B.}~\bibnamefont {Malkin}}, \bibinfo {author}
  {\bibfnamefont {A.}~\bibnamefont {Stepanov}}, \ and\ \bibinfo {author}
  {\bibfnamefont {B.}~\bibnamefont {Barbara}},\ }\href {\doibase
  10.1038/nnano.2006.174} {\bibfield  {journal} {\bibinfo  {journal} {Nature
  Nanotech.}\ }\textbf {\bibinfo {volume} {2}},\ \bibinfo {pages} {39}
  (\bibinfo {year} {2007})}\BibitemShut {NoStop}%
\end{thebibliography}
\end{document}